\documentclass{article}
\usepackage{locata,amsmath,graphicx,url,times}
\usepackage[T1]{fontenc}
\usepackage{epstopdf}
\usepackage{amssymb,amsmath,amsthm}
\usepackage{xifthen}
\usepackage{epstopdf}
\usepackage{xcolor}
\def\UrlBreaks{\do\/\do-}

\newcommand{\ing}[1]{\mathsf{#1}}
\newcommand{\Rn}[1]{\ifthenelse{\equal{#1}{}}{\mathbb{R}}{\mathbb{R}^{\ing{#1}}}}
\newcommand{\Cn}[1]{\mathbb{C}^{\ing{#1}}}
\newcommand{\Rset}[2]{\ifthenelse{\equal{#2}{1}}{\in \Rn{\ing{#1}}}{\in \Rn{\ing{#1} \times \ing{#2}}}}
\newcommand{\Cset}[2]{\ifthenelse{\equal{#2}{1}}{\in \Cn{\ing{#1}}}{\in \Cn{\ing{#1} \times \ing{#2}}}}
\newcommand{\vect}[1]{\boldsymbol{\mathbf{\MakeLowercase{#1}}}}
\newcommand{\mtrx}[1]{\boldsymbol{\mathbf{\MakeUppercase{#1}}}}

\newcommand{\transp}[1]{#1^{\mathsf{T}}}
\newcommand{\htransp}[1]{#1^{\mathsf{H}}}

\newcommand{\ind}[2]{\ifthenelse{\equal{#2}{}}{\chi_{#1}}{\chi_{#1}\left( #2 \right)}}

\newcommand{\iter}[2]{#1^{\ing{(#2)}}}

\definecolor{monvert}{rgb}{0,0.8,0.5}

\title{TRAMP: Tracking by a Realtime AMbisonic-based Particle filter}

\twoauthors
  {Sr\dj{}an Kiti\'c}
    {Orange Labs\\
4 Rue du Clos Courtel\\
     35510 Cesson-S\'evign\'e, France \\
     srdan.kitic@orange.com}
  {Alexandre Gu\'erin}
    {Orange Labs\\
4 Rue du Clos Courtel\\
     35510 Cesson-S\'evign\'e, France \\
     alexandre.guerin@orange.com}

\begin{document}

\ninept
\maketitle

\begin{sloppy}

\begin{abstract}
This article presents a multiple sound source localization and tracking system, fed by the Eigenmike array. The First Order Ambisonics (FOA) format is used to build a pseudointensity-based spherical histogram, from which the source position estimates are deduced. These instantaneous estimates are processed by a well-known tracking system relying on a set of particle filters. While the novelty within localization and tracking is incremental, the fully-functional, complete and real-time running system based on these algorithms is proposed for the first time. As such, it could serve as an additional baseline method of the LOCATA challenge. 
\end{abstract}


\section{Introduction}
\label{sec:intro}

Recent commercial success of the so-called smart speakers \cite{mreversal} has  amplified the need for multichannel signal enhancement tools in order to cope with noisy environments. Among them, sound source localization by itself is a challenging problem that has been considered for long as a known prior \cite{vincent_audio_2018,virtanen_techniques_2012} and not been fully solved to date, despite long-standing research efforts (\emph{e.g.} \cite{sheng2005maximum,pavlidi2013real,dibiase2001robust,kitic2013review,jarrett20103d}). Accompanied by the practical requirement for a simultaneous tracking (\emph{i.e.} assignement of pertinent labels to the detected sources), the task becomes considerably more difficult and demands elaborate algorithmic frameworks, \emph{e.g.} \cite{valin2007robust, grondin2013manyears, evers2017source, fallon2012acoustic, ward2003particle}. 

Despite the evident prevalence of the problem, the recently announced IEEE-AASP Challenge on Acoustic Source Localization and Tracking (LOCATA) \cite{lollmann2018locata} is amongst first comprehensive benchmarks organized by the community. For this reason, we believe it is important to include different baseline methods in evaluation, and thus drop the anchor for future research. The challenge organizers have provided one such reference, based on MUltiple SIgnal Classification (MUSIC) algorithm for localization, and an ensemble of Kalman filters for tracking \cite{lollmann2018locata}. In this work, we propose another baseline, adapted to the Eigenmike spherical array and exploiting the FOA (\emph{i.e. B-format} \cite{gerzon1992ambisonic}) signals, termed \emph{Tracking by a Real-time AMbisonic-based Particle filter (TRAMP)}. Position estimates are yielded from the pseudointensity histogram \cite{jarrett20103d,tervo2009direction}, while the tracking is performed using the particle filtering method of Valin et al. \cite{valin2007robust}. Nevertheless, some modifications of the original approaches are proposed, as discussed in the following sections.

The paper is organized as follows: in Section~\ref{sec:localization} we present the ``instantaneous'' localization 
algorithm, while Section~\ref{sec:tracking} discusses the particle filtering approach used for detecting and tracking sound sources. Section~\ref{sec:results} presents the results obtained on the LOCATA developement dataset, and Section~\ref{sec:conclusion} concludes the article.

\section{Localization}
\label{sec:localization}

\subsection{FOA format and pseudointensity}

The (Higher Order) Ambisonics (HOA) has gained considerable attention as an attractive, capture/reproduction system-independent, spatial representation of a sound field \cite{gerzon1992ambisonic,daniel2003further,nicol2010sound}. Multichannel spherical microphone signals are converted into so-called \emph{HOA components} by computing weighted scalar products between the measurements and spherical harmonics \cite{nicol2010sound}. Within certain vicinity from the spherical microphone\footnote{The largest source-free sphere centered on the microphone array \cite{nicol2010sound}.}, the sound field can be reproduced, with high accuracy, from a linear combination of the infinite number of HOA components. In practice, the HOA representation is truncated (one keeps only a limited number of components), which further limits the sound field volume that can be reproduced with sufficient accuracy, but, at the same time, saves computational resources.

If the pseudointensity vectors are used for localization, it is sufficient to extract only HOA components of order one, \emph{i.e.} FOA. FOA format consists of four channels: $\vect{w} = \left[ w(\ing{t}) \right]_{\ing{t}=1\hdots N}$, $\vect{x} = \left[ x(\ing{t}) \right]_{\ing{t}=1\hdots N}$, $\vect{y} = \left[ y(\ing{t}) \right]_{\ing{t}=1\hdots \ing{N}}$ and $\vect{z} = \left[ z(\ing{t}) \right]_{\ing{t}=1\hdots N}$, where $\ing{t}$ denotes the time sample. The conversion is a linear operation, that can be compactly represented in matrix form as 
\begin{equation}
	\transp{\left[ \begin{matrix} \vect{w} \; \vect{x} \; \vect{y} \; \vect{z} \end{matrix} \right]} = \mtrx{E}_1 \htransp{\mtrx{Y}}_1 \mtrx{M},
\end{equation}
where $\mtrx{Y}_1$ is the matrix of up-to-first-order spherical harmonics, $\mtrx{E}_1$ is the equalization matrix (possibly a matrix of filters \cite{craven77}), and $\mtrx{M}$ is the matrix of vertically-stacked raw multichannel recordings. Usually, the matrices $\mtrx{Y}_1$ and $\mtrx{E}_1$ are precomputed for a given spherical array.

The FOA channels have immediate physical interpretation. The omnidirectional component $w(\ing{t})$ amounts to the acoustic pressure averaged over all microphone channels $p_{\ing{i}}(\ing{t}), \ing{i}=1\hdots\ing{M}$, while $x(\ing{t})$, $y(\ing{t})$ and $z(\ing{t})$ approximate spatial gradients oriented along corresponding Cartesian coordinate axes. The linearized fluid momentum equation relates the spatial gradient vector $\nabla p(\ing{t})$ and particle velocity $\vect{u}(\ing{t})$ (\emph{cf.} \cite[eq.2.29]{merimaa2006analysis}), enabling the approximation of the \emph{sound intensity} vector $\mtrx{I}(\ing{t})$ (up to a multiplicative coefficient):
\begin{equation}
	\mtrx{I}(\ing{t}) = p(\ing{t})\vect{u}(t) \propto p(\ing{t})\nabla \vect{p}(t) \approx  w(\ing{t}) \left[ \begin{matrix} x(\ing{t}) \\ y(\ing{t}) \\ z(\ing{t}) \end{matrix} \right]
\end{equation}

For narrowband signals, one can define \emph{active intensity} \cite{jacobsen1991note}:
\begin{equation}
	\mtrx{I}_a(\omega) \propto  \Re \left( p(\omega)^* \nabla p(\omega) \right) \approx \Re \left( W(\omega)^* \left[ \begin{matrix} X(\omega) \\ Y(\omega) \\ Z(\omega) \end{matrix} \right] \right) := \hat{\mtrx{I}}(\omega),
\end{equation}
where $\Re(\cdot)$ denotes the real part of an argument, $\omega$ is angular frequency, and $p(\omega)$, $W(\omega)$, $X(\omega)$, $Y(\omega)$ and $Z(\omega)$ are the Fourier representations of the corresponding quantities. The estimate $\hat{\mtrx{I}}(\omega)$ is termed ``pseudointensity'' \cite{jarrett20103d}. 

In practice, in order to account for the dynamics of the acoustic scene and the sources' non-stationarity, $\hat{\mtrx{I}}(\ing{t},\omega)$ is computed online from the Short Time Fourier Transform (STFT) of the FOA channels (we use the $0.04s$ frame with $50\%$ overlap). Another point is the frequency range validity of HOA representation that depends on the order, the microphone configuration and the type of encoding  \cite{daniel2003further}. For FOA deduced from an Eigenmike, we restricted the frequency range to ${[400,7000]\; \text{Hz}}$. 

\subsection{Vocal Activity Detection}
The localisation task described in this paper is triggered by a frame-based Voice Activity Detection (VAD). This VAD is derived from the Hendriks algorithm \cite{gerkmann2012unbiased}, designed to quickly track the noise power when merged with speech. The Hendriks algorithm is based on the observation that speech is intermittent, whereas noise is statistically more regular over time. Using complex Gaussian models for speech and noise spectral components coupled with a speech presence probability estimation, the method computes, in each frequency band, an unbiased-MMSE noise power estimate $\hat{\sigma}^{2}_{n}(\ing{t},\omega)$. We apply this algorithm to the $W(\ing{t},\omega)$ channel in order to produce estimates of the time-frequency \emph{a posteriori} SNRs $\hat{\gamma}(\ing{t},\omega) = W^{2}(\ing{t},\omega) / \hat{\sigma}^{2}_{n}(\ing{t},\omega) - 1$. A frame-SNR $\hat{\gamma}(\ing{t})$ is deduced by integrating $\hat{\gamma}(\ing{t},\omega)$ over frequency: $\hat{\gamma}(\ing{t}) = \int_{400.2\pi}^{7000.2\pi} \hat{\gamma}(\ing{t},\omega) d\omega$. The binary frame-VAD is finally obtained by comparing the frame-SNR $\hat{\gamma}(\ing{t})$ to a fixed threshold (here 7dB)\footnote{Hereafter, unless essential, we drop the dependency on $\ing{t}$ and $\omega$.}.

\subsection{Direction-Of-Arrival (DOA) estimation}

Pseudointensity characterizes the flow of energy and, in anechoic/single-source setting, points towards the DOA of a sound source \cite{pulkki_spatial_2007}. In the more realistic, reverberant and multiple-source case, the DOA estimation is more problematic \cite{pulkki_parametric_2017}. Ideally, we should consider only those frequency bands that are occupied by one source, and are less affected by reverberation, \emph{i.e.} which behave as plane waves in the far field. The FOA components of a plane wave source at azimuth $\theta$ and elevation $\varphi$ admit simple expressions \cite[p.38]{merimaa2006analysis}:
\begin{align}\label{eqPlaneWave}
	W = p, \quad X & = \sqrt{C} p \cos \theta \cos \varphi, \\
	Y = \sqrt{C} p \sin \theta \cos \varphi, & \quad Z = \sqrt{C} p \sin \varphi, \nonumber
\end{align}
where the constant $C$ depends on the FOA encoding format and is considered known in advance. Hence, the plane wave FOA components are linked by the deterministic relation:
\begin{equation}\label{eqPlaneWaveR}
	R := \frac{X^2+Y^2+Z^2}{W^2} = C.
\end{equation}

When facing reverberation, the plane wave hypothesis is no longer exact and the equations \eqref{eqPlaneWave} and \eqref{eqPlaneWaveR} appear noisy. Hence, we propose to use statistic properties of the pseudointensity vector to identify the source locations. Firstly, 3D-space is discretized on the Lebedev grid \cite{lebedev1999quadrature} of $\ing{N}=974$ nodes identified by their spherical coordinates $(\theta, \varphi)$. Then, the histogram of time-frequency DOAs $\angle \hat{\mtrx{I}}(\ing{t},\omega)$ is built, where the value of the histogram bin $(\theta,\varphi)$ at frame $\ing{t}$ is given by\footnote{For readability reasons, we abuse notation and associate the time index $\ing{t}$ to the STFT frame index.}:
\begin{equation}
	b_{\theta,\varphi}(\ing{t}) := \sum\limits_{\ing{t}' = \ing{t} - \ing{T}}^{\ing{t}} \sum\limits_{\omega \in \Omega } \frac{\max(\hat{\gamma}(\ing{t}',\omega), 0)}{\left( 1 + |C - R(\ing{t}',\omega)| \right)^2},
\end{equation}
where  $\Omega = \{ \omega' \mid (\theta,\varphi) \leftarrow \angle \hat{\mtrx{I}}(\ing{t}',\omega') \}$, and ``$\leftarrow$'' is to be understood as $\angle \hat{\mtrx{I}}(\ing{t}',\omega')$ being quantized to $(\theta,\varphi)$. This aggregates the directions falling into the same bin across frequencies and a time window of certain size, and puts larger weight to those having high SNR and are respectful to the criterion \eqref{eqPlaneWaveR}. A large number of frames $\ing{T}$ produces a smoother histogram, but affects the adaptivity/response time: in practise, $\ing{T}$ is set to correspond to a $1$~s buffer. 

When the VAD triggers a speech presence, the histogram values are linearly expanded to the interval $[0 ; 1]$, and the bins above $0.3$ are selected. Next, a nearest-neighbour (in the sense of angular distance between the grid nodes) Gaussian filter of variance $0.2$ and support size $50$, is applied to the selected bins. The set of potential DOAs at time $\ing{t}$, $\iter{o}{t} :=  \{ \iter{o}{t}_{\ing{q}} \}_{0\hdots\ing{Q}-1} = \{(\theta_{\ing{q}},\phi_{\ing{q}})\}_{0\hdots\ing{Q}-1}$, $\ing{Q} \geq 0$, is the subset of these directions whose filtered histogram values $P_{\ing{q}}$ are local maxima in their respective neighbourhoods.

\section{Tracking}
\label{sec:tracking}

Due to reverberation, the instantaneous DOA observations $\iter{o}{t}$ are noisy and may contain some outliers also referred as false alarms (FAs). Experience shows that their associated $P_{\ing{q}}$ values may be roughly interpreted as prior probability of $(\theta_{\ing{q}},\phi_{\ing{q}})$ \emph{not} being FAs. Hence, we feed this information into a particle-based tracker (largely founded on \cite{valin2007robust,grondin2013manyears}) that identifies sources and yields their DOA estimates. The remainder of this section describes recursive Bayesian estimation of the involved quantities.

\subsection{State-space model}

Let, at the time $\ing{t}$, be $\ing{S}$ tracked sources. Each tracked source $\ing{s}$ contains $\ing{P} = 300$ Langevin dynamical models \cite{ward2003particle} (``particles''):
\begin{align}\label{eqLangevin}
	\iter{\vect{v}}{t}_{\ing{s},\ing{p}} & = a_{\ing{p}} \iter{\vect{v}}{t-1}_{\ing{s},\ing{p}} + b_{\ing{p}} \vect{n} \\
	\iter{\vect{p}}{t}_{\ing{s},\ing{p}} & = \iter{\vect{p}}{t-1}_{\ing{s},\ing{p}} + \Delta T \iter{\vect{v}}{t}_{\ing{s},\ing{p}}, \; \; \ing{p} \in [1,\ing{P}], \nonumber
\end{align}
where $\iter{\vect{p}}{t}_{\ing{s},\ing{p}}$ is the position vector in Cartesian coordinates of the $\ing{p}^{th}$ particle associated to the $\ing{s}^{th}$ source, $\iter{\vect{v}}{t}_{\ing{s},\ing{p}}$ is the velocity, $\vect{n}$ is a unit-variance centered white Gaussian noise, $\Delta T$ is the frame rhythm (in ms), while the fixed parameters $a_{\ing{p}}$ and $b_{\ing{p}}$ are adopted from \cite{grondin2013manyears}. At every iteration, and for all existing particles, the forward pass \eqref{eqLangevin} is executed first.

Particles bear importance probabilities, \emph{i.e. weights}
\begin{equation}
	\iter{\mathrm{w}}{t}_{\ing{s},\ing{p}} = P(\iter{\vect{p}}{t}_{\ing{s},\ing{p}} \mid \iter{\vect{o}}{t}), \quad  \iter{\vect{o}}{t} = \{ \iter{o}{t}, \iter{o}{t-1}, \hdots \iter{o}{0} \},
\end{equation}
used to estimate the current position and velocity of a source
\begin{equation}\label{eqPosition}
	\iter{\vect{p}}{t}_{\ing{s}} = \sum\limits_{\ing{p}} \iter{\mathrm{w}}{t}_{\ing{s},\ing{p}} \iter{\vect{p}}{t}_{\ing{s},\ing{p}}, \quad \iter{\vect{v}}{t}_{\ing{s}} = \sum\limits_{\ing{p}} \iter{\mathrm{w}}{t}_{\ing{s},\ing{p}} \iter{\vect{v}}{t}_{\ing{s},\ing{p}},
\end{equation}
as well as the probability density of the event that observation $\iter{o}{t}_{\ing{q}}$ originates from the source $\ing{s}$:
\begin{align}\label{eqAssociation}
	p( \iter{o}{t}_{\ing{q}} \mid \ing{s} ) & = \sum\limits_{\ing{p}} \iter{\mathrm{w}}{t-1}_{\ing{s},\ing{p}} p( \iter{o}{t}_{\ing{q}} \mid \iter{\vect{p}}{t}_{\ing{s},\ing{p}} ), \; \text{with} \\
	p( \iter{o}{t}_{\ing{q}} \mid \iter{\vect{p}}{t}_{\ing{s},\ing{p}} ) &= \mathcal{N}\left(\iter{\vect{p}}{t}_{\ing{s},\ing{p}}, \frac{0.008}{1 + 0.2\angle(\iter{\vect{v}}{t}_{\ing{s},\ing{p}}, \iter{o}{t}_{\ing{q}}-\iter{\vect{p}}{t-1}_{\ing{s},\ing{p}})}  \right).
\end{align}
The intuition behind adaptive variance is that the probability should be high if  $\iter{o}{t}_{\ing{q}}$ is very close to the new prediction $\iter{\vect{p}}{t}_{\ing{s},\ing{p}}$, and - of lower importance - if the displacement direction matches the particle direction of the velocity vector. 

\subsection{Weight estimation and creation/deletion of the sources} 

Let us define association functions $f_{\ing{m}}(\ing{q})$ of observations to sources:
\begin{equation}
	f_{\ing{m}}(\ing{q}) = \begin{cases}
					-2, & \mathcal{H}_{\text{FA}}\colon \text{$\iter{o}{t}_{\ing{q}}$ is a FA.} \\
					-1, & \mathcal{H}_{\text{new}}\colon \text{$\iter{o}{t}_{\ing{q}}$ is a new source.} \\
					\ing{s}, & \mathcal{H}_{\ing{s}}\colon \text{$\iter{o}{t}_{\ing{q}}$ comes from source $\ing{s}$.}
				 \end{cases}
\end{equation}
For a number of Q observations and S sources, the cardinal of the set of association functions $ f_{\ing{m}}(\ing{q}) $ is $(\ing{S}+2)^{\ing{Q}}$, and we assume that one of them is the correct mapping of observations to sources/FAs. In reality, one should also consider the cases where two or more sources ``share'' the same observation (for instance when 2 sources cross), but the cardinality of such function space becomes much larger. Moreover, the mappings are to be exploited in the probabilistic sense: each source $\ing{s}$ is affected by all observations, and the fact that a "shared" observation would not have been associated to a source should not impact too deeply its position/velocity predictions.

Let $P(f_{\ing{m}} \mid \iter{o}{t})$ represent the conditional probability of $f_{\ing{m}}$ being the correct mapping, given the current set of observations $\iter{o}{t}$. Having evaluated all such probabilities, one can calculate the marginals of the hypotheses $\mathcal{H}_{\text{FA}}$, $\mathcal{H}_{\text{new}}$ and $\mathcal{H}_{\ing{s}}$, for each $\iter{o}{t}_{\ing{q}}$:
\begin{align}
	\iter{P}{t}_{\ing{q}}(\mathcal{H}_{\text{FA}}) &= \sum\limits_{\ing{m}} \delta_{-2,f_{\ing{m}}(\ing{q})} P(f_{\ing{m}} \mid \iter{o}{t}), \\
	\iter{P}{t}_{\ing{q}}(\mathcal{H}_{\text{new}}) &= \sum\limits_{\ing{m}} \delta_{-1,f_{\ing{m}}(\ing{q})} P(f_{\ing{m}} \mid \iter{o}{t}),  \\
	\iter{P}{t}_{\ing{q}}(\mathcal{H}_{\ing{s}}) &= \sum\limits_{\ing{m}} \delta_{\ing{s},f_{\ing{m}}(\ing{q})} P(f_{\ing{m}} \mid \iter{o}{t}),
\end{align}
where $\delta_{\cdot,\cdot}$ denotes the Kronecker delta function. Observe that $\iter{P}{t}_{\ing{q}}(\mathcal{H}_{\text{FA}}) + \iter{P}{t}_{\ing{q}}(\mathcal{H}_{\text{new}}) + \sum_{\ing{s}} \iter{P}{t}_{\ing{q}}(\mathcal{H}_{\ing{s}}) = \sum_{\ing{m}} P(f_{\ing{m}} \mid \iter{o}{t}) = 1$.

The probability that the source $\ing{s}$ \emph{has been} observed at time $\ing{t}$ is
\begin{equation}
	\iter{P}{t}_{\ing{s}} = \frac{1}{Q} \sum\limits_{q=0}^{Q-1} \iter{P}{t}_{\ing{q}}(\mathcal{H}_{\ing{s}}),
\end{equation}
if $\ing{Q}>0$; otherwise, $\iter{P}{t}_{\ing{s}} = 0$. If $\iter{P}{t}_{\ing{s}} \geq 0.3$, the source $\ing{s}$ is marked as enabled, and disabled otherwise. At a given time instance, we produce the output estimates only for the sources that have been enabled for longer than $0.1$ s: this simple hangover rule greatly reduces spurious detections. On the other hand, active sources are effectively deleted if they have been disabled for more than $0.2$ s consecutively (the other sources may also be tracked, but are not ``visible''). 

Whenever $\iter{P}{t}_{\ing{q}}(\mathcal{H}_{\text{new}}) \geq 0.8$, the observation $\iter{o}{t}_{\ing{q}}$ is declared to be a new source ($\ing{S}\gets\ing{S}+1$). It is initialized by setting its particle positions $\iter{\vect{p}}{t}_{\ing{s},\ing{p}}$ equal to Cartesian coordinates\footnote{Converted assuming the arbitrary, but fixed radius $r>0$.} of $(\theta_{\ing{q}},\phi_{\ing{q}})$, their velocities to $\iter{\vect{v}}{t}_{\ing{s},\ing{p}} = \vect{0}$, and letting ${\iter{P}{t}_{\ing{S}} = \iter{P}{t}_{\ing{q}}(\mathcal{H}_{\text{new}})}$. After the (potential) addition of new sources, if there is a limit on the total number of sources simultaneously tracked $\ing{S}_{\max}$, we suppress the excess ones by keeping only those with largest probabilities $\iter{P}{t}_{\ing{s}}$.

Particle weight derivations are straightforward, but somewhat tedious, thus we provide the resulting expressions only (the interested reader may consult \cite{valin2007robust} for the detailed procedure):
\begin{align}
	\iter{\mathrm{w}}{t}_{\ing{s},\ing{p}} &= \frac{p(\iter{\vect{p}}{t}_{\ing{s},\ing{p}} \mid \iter{o}{t}) \iter{\mathrm{w}}{t-1}_{\ing{s},\ing{p}}}{\sum\limits_{\ing{p}} p(\iter{\vect{p}}{t}_{\ing{s},\ing{p}} \mid \iter{o}{t}) \iter{\mathrm{w}}{t-1}_{\ing{s},\ing{p}}}, \; \text{where}\\
	p(\iter{\vect{p}}{t}_{\ing{s},\ing{p}} \mid \iter{o}{t}) &= \frac{1}{\ing{P}}(1-\iter{P}{t}_{\ing{s}}) + \iter{P}{t}_{\ing{s}} \frac{ \sum\limits_{\ing{q}}  \iter{P}{t}_{\ing{q}}(\mathcal{H}_{\ing{s}}) p( \iter{o}{t}_{\ing{q}} \mid \iter{\vect{p}}{t}_{\ing{s},\ing{p}} )}{ \sum\limits_{\ing{p}}\sum\limits_{\ing{q}}  \iter{P}{t}_{\ing{q}}(\mathcal{H}_{\ing{s}}) p( \iter{o}{t}_{\ing{q}} \mid \iter{\vect{p}}{t}_{\ing{s},\ing{p}} ) }.
\end{align}

The well-known drawback of the method is the exponential increase of estimate variance with $\ing{t}$ \cite{doucet2009tutorial}. This is addressed by resampling, \emph{i.e.} occasional particle set replacement, through sampling from the probability mass function defined by $\{ \iter{\mathrm{w}}{t}_{\ing{s},\ing{p}} \}_{\ing{p}}$. For any source $\ing{s}$, the resampling trigger condition is
\begin{equation}
	\left( \sum_{\ing{p}} (\iter{\mathrm{w}}{t}_{\ing{s},\ing{p}})^2 \right)^{-1} < 0.7 \ing{P}.
\end{equation}
Afterwards, weights of the new particles are uniformly set to $1/\ing{P}$.

\subsection{Association function probabilities}

Applying Bayes' rule to $P(f_{\ing{m}} \mid \iter{o}{t})$ gives 
\begin{equation}
	P(f_{\ing{m}} \mid \iter{o}{t}) \propto p(\iter{o}{t} \mid f_{\ing{m}})P(f_{\ing{m}}). 
\end{equation}
Under the aforementioned ``correctness'' assumption, computing the right-hand side of the proportionality relation for all $f_{\ing{m}}$, and then normalizing the results, yields the probabilities $P(f_{\ing{m}} \mid \iter{o}{t})$. 

The probability densities $p(\iter{o}{t} \mid f_{\ing{m}})$ are estimated by assuming conditional independence of observations given the mapping function: $p(\iter{o}{t} \mid f_{\ing{m}}) = \prod_{\ing{q}} p( \iter{o}{t}_{\ing{q}} \mid f_{\ing{m}}(\ing{q}) )$, where we set $p( \iter{o}{t}_{\ing{q}} \mid -2 ) = p( \iter{o}{t}_{\ing{q}} \mid -1 ) = (4\pi)^{-1}$, or use \eqref{eqAssociation}, otherwise. Analogously, the prior probabilities of correct assignment of each observation are considered independent, thus we have $P(f_{\ing{m}}) = \prod_{\ing{q}} P(f_{\ing{m}}(\ing{q}) )$, where
\begin{equation}
	P(f_{\ing{m}}(\ing{q}) ) = 
	\begin{cases}
		0.5(1 - P_{\ing{q}}), & f_{\ing{m}}(\ing{q}) = -2,\\
		0.05P_{\ing{q}}, & f_{\ing{m}}(\ing{q}) = -1,\\
		P_{\ing{q}} \iter{P}{t}_{\text{obs}}(f_{\ing{m}}(\ing{q}) \mid \iter{\vect{o}}{t-1}), & \text{otherwise}.
	\end{cases}
\end{equation}

${\iter{P}{t}_{\text{obs}}(\ing{s} \mid \iter{\vect{o}}{t-1})}$ is the probability of the source $\ing{s}$ being \emph{observable} at time $\ing{t}$, and equals the product of ${\iter{P}{t}_{\text{exist}}(\ing{s} \mid \iter{\vect{o}}{t-1})}$ (probability that the source exists), and ${\iter{P}{t}_{\text{act}}(\ing{s} \mid \iter{\vect{o}}{t-1})}$ (probability of the source being active). Due to spatial constraints, we write down the final expressions only, and again refer the reader to the article \cite{valin2007robust}:
\begin{align}
	& \iter{P}{t}_{\text{exist}}(\ing{s} \mid \iter{\vect{o}}{t-1}) = \iter{P}{t-1}_{\ing{s}} + (1-\iter{P}{t-1}_{\ing{s}}) \frac{0.5 \iter{P}{t-1}_{\text{exist}}(\ing{s} \mid \iter{\vect{o}}{t-2} )}{1 - 0.5 \iter{P}{t-1}_{\text{exist}}(\ing{s} \mid \iter{\vect{o}}{t-2} )}, \nonumber \\
	& \iter{P}{t}_{\text{act}}(\ing{s} \mid \iter{\vect{o}}{t-1}) = 0.4\iter{P}{t-1}_{\text{act}}(\ing{s} \mid \iter{\vect{o}}{t-1}) + 0.3, \; \text{with} \\
	& \iter{P}{t-1}_{\text{act}}(\ing{s} \mid \iter{\vect{o}}{t-1}) = \nonumber \\
	& \left(1 + \frac{ (1 - \iter{P}{t-1}_{\text{act}}(\ing{s} \mid \iter{\vect{o}}{t-2})) (1-\iter{P}{t-1}_{\text{act}}(\ing{s} \mid \iter{o}{t-2})) }{\iter{P}{t-1}_{\text{act}}(\ing{s} \mid \iter{\vect{o}}{t-2}) \iter{P}{t-1}_{\text{act}}(\ing{s} \mid \iter{o}{t-2})} \right)^{-1}, \; \text{and} \nonumber \\
	& \iter{P}{t-1}_{\text{act}}(\ing{s} \mid \iter{o}{t-2}) = 0.15 + 0.85\iter{P}{t-1}_{\ing{s}}.  \nonumber
\end{align}

As the final step of each tracker function call, we evaluate angular distances between the estimated positions \eqref{eqPosition} of all tracked sources. The closest \emph{pairs} are isolated, and for these we check whether the distances are lower than $5^{\circ}$. Within each such pair, we choose the source that has been enabled for a shorter time, and decrease its probability ${\iter{P}{t}_{\text{exist}}(\ing{s} \mid \iter{\vect{o}}{t-1})}$ by a factor of $0.95$. This heuristics alleviates the issue of redundant sources, but does not prevent occasional trajectory crossings. However, it limits the ability of tracking fixed, but very closely positioned sources.

\section{Results}
\label{sec:results}

We exercise the method on the developement dataset of the \mbox{LOCATA} challenge \cite{lollmann2018locata}. Particularly, we consider the tasks involving the static Eigenmike array, as it mimics the conditions of use of a smart speaker:
\begin{enumerate}
	\item Tracking of a single, stationary loudspeaker,
	\item Tracking of multiple, stationary loudspeakers,
	\item Tracking of a single, moving talker, and,
	\item Tracking of multiple, moving talkers.
\end{enumerate}

Before discussing the evaluation metric and results, let us remark that the presented algorithms were not pre-parametrized for any specific task (\emph{e.g.}, in the tasks $1$ and $3$, they are unaware of being in the single-source scenario). While such additional information could significantly improve the overall performance, it is rarely available in real-life situations. However, we limit the number of instantaneous observations\footnote{By keeping at most $\ing{Q}_{\max}$ observations $\{\iter{o}{t}_{\ing{q}}\}$ having the largest $P_{\ing{q}}$.} $\ing{Q}_{\max}=4$, and the number of simultaneous trajectories $\ing{S}_{\max}=4$, in order to ensure the real-time execution of the present, single-thread Python implementation. Note that the method should gracefully parallelize, since the most expensive step -- evaluation of $p( \iter{o}{t}_{\ing{q}} \mid \iter{\vect{p}}{t}_{\ing{s},\ing{p}} )$ -- can be performed independently for each particle $\ing{p}$ (which is not exploited in the current version).

The choice of evaluation metric can significantly affect the obtained results. In case of multisource tracking, an appropriate metric should take into account not only localization accuracy, but also False Alarm Rate (FAR), False Negative Rate (FNR), and consistency/continuity of track IDs. As suggested in \cite{locataEvalaution}, the Optimal Subpattern Assignment (OSPA) distance \cite{ristic2011metric} would be a proper way to measure performance. Nevertheless, the results of the MUSIC/Kalman baseline algorithm \cite{lollmann2018locata} are provided for a different metric, which we adopt here, in order to have comparable scores. Hence, as proposed in \cite{lollmann2018locata}, the time-averaged azimuth error is computed for all pairs of ground-truth source trajectories and estimated source tracks. Then, the Hungarian algorihtm \cite{kuhn1955hungarian}, based on the resulting cost matrix, is used to find the optimal assignment between ground-truth and estimated tracks (note that the choice of time-averaged metric favors tracks of short duration, which may represent redundant sources). The overall error per recording is the average azimuth error for the selected pairs, and the final per-task error is given by computing the average and standard deviation of per-recording errors over all recordings for a given task. 

\begin{table}
\centering 
\begin{tabular}{|c||c|c|}
	\hline
	Task & Mean & Std \\
	\hline 
	1 & 6.19 & 1.18 \\ \hline
	2 & 6.88 & 5.23 \\ \hline
	3 & 15.05 & 3.09 \\ \hline
	4 & 11.97 & 7.04 \\ 
	\hline
\end{tabular}
\quad
\begin{tabular}{|c||c|c|}
	\hline
	Task & Mean & Std \\
	\hline 
	1 & 6.02 & 5.19 \\ \hline
	2 & 4.50 & 2.39 \\ \hline
	3 & 6.62 & 1.74\\ \hline
	4 & 7.21 & 1.25 \\ 
	\hline
\end{tabular}

\caption{Azimuth (left) and elevation (right) errors, in degrees.} \label{tabResults}
\end{table}

The results of the proposed baseline method, using the discussed azimuth metric for the track assignment, are given in Table~\ref{tabResults}. The obtained mean azimuth errors are lower than the ones provided in \cite{lollmann2018locata} for the Eigenmike array. We have to again remind the reader that these results should be taken with a grain of salt, due to the suboptimal choice of track assignment evaluation metric. One notable artefact of this choice is the smaller azimuth error on Task~4 compared to the azimuth error on Task~3, despite the former being a markedly more difficult problem. The more accurate and comprehensive benchmark (including the FAR, FNR etc. results), obtained on the larger evaluation dataset, should become available upon completion of the LOCATA challenge.

\section{Conclusion}
\label{sec:conclusion}

We proposed TRAMP, a multisource localization and tracking algorithm, based on the sound intensity estimation coupled with an adaptation of the well-known particle filtering algorithm. By exploiting the Ambisonics format, the method is essentially agnostic to the configuration of the microphone array, as long as the FOA conversion is done properly. The initial results indicate that the TRAMP method performs favourably against the MUSIC/Kalman baseline. Further improvements may include more accurate localization methods, \emph{e.g.} by exploiting the HOA signals \cite{hafezi2017augmented} or neural networks \cite{adavanne2017direction,perotin2018crnn}, and/or more sophisticated and efficient tracking schemes.


\bibliographystyle{IEEEtran}
\bibliography{refs18}
%
%
%
%
%
%
%
%
%

\end{sloppy}
\end{document}